\documentclass[sigconf,9pt]{acmart}

\usepackage{amsmath,amsfonts}

\usepackage{amssymb}

\usepackage{pgfplots,wrapfig}

\copyrightyear{2026}
\acmYear{2026}
\setcopyright{cc}
\setcctype{by}
\acmConference[DAC '26]{63rd ACM/IEEE Design Automation Conference}{July 26--29, 2026}{Long Beach, CA, USA}
\acmBooktitle{63rd ACM/IEEE Design Automation Conference (DAC '26), July 26--29, 2026, Long Beach, CA, USA}
\acmDOI{10.1145/3770743.3804400}
\acmISBN{979-8-4007-2254-7/2026/07}

\usepackage{float}

\usepackage{graphicx}
\usepackage[table]{xcolor}
\usepackage[normalem]{ulem}
\usepackage{lipsum}

\usepackage{booktabs}
\usepackage{multirow}
\usepackage{tabularx}
\usepackage{tablefootnote}

\makeatletter
\renewcommand{\linespread}[1]{}
\makeatother
\AtBeginDocument{\setlength{\baselineskip}{\dimexpr\baselineskip*935/1000\relax}}

\usepackage{tikz}
\newcommand{\circled}[1]{\tikz[baseline=(char.base)]{
  \node[shape=circle,draw=black,fill=black,text=white,inner sep=0.5pt] (char) {#1};}}

\usepackage[ruled,lined,linesnumbered,commentsnumbered,noend]{algorithm2e}
\SetKwComment{Comment}{/* }{ */}

\newcommand{\xmark}{\(\times\)}
\usepackage{cancel}
\usepackage{array,enumitem,soul,pifont}

\newcommand{\detcwe}[1]{%
  {\begingroup
   \setlength{\fboxsep}{0.5pt}
   \colorbox{green!20}{\textbf{#1}}%
   \endgroup
  }%
}

\newcommand{\atlaslegendcell}{%
  \multicolumn{5}{@{}l@{}|}{%
    \multirow{4}{*}{%
      \begingroup
      \scriptsize
      \setlength{\fboxsep}{1pt}%
      \colorbox{gray!07}{%
        \parbox[t]{0.35\linewidth}{%
          \centering\textbf{How to Read the Table}\par
          \raggedright CWE formatting:\par
          \begin{itemize}[leftmargin=*, itemsep=1pt, topsep=1pt]
            \vspace{3pt}
            \item \detcwe{Green} = Correct;
            \item \st{Strikethrough} = Undetected by \emph{ATLAS};
            \item {\color{blue}Blue Texts} = Assumed correct (no baseline for 5 CWEs).
            \item Unbolded CWEs = Relevant CWEs (Sec.~\ref{subsubsec:relevant_cwes})
            \item \checkmark/\xmark\ Indicate pass/fail (Prop.\ and FV).
            \vspace{3pt}
          \end{itemize}%
        }%
      }%
      \endgroup
    }%
  }%
}
\newcommand{\atlaslegendempty}{\multicolumn{5}{@{}l@{}|}{}}

\usepackage{textcomp}
\usepackage{subcaption}
\usepackage{comment}

\providecommand{\IEEEauthorblockA}[1]{#1}

\title{ATLAS: AI-Assisted Threat-to-Assertion Learning for System-on-Chip Security Verification}

\author{\small Ishraq Tashdid$^{1*}$, Kimia Tasnia$^{1*}$, Alexander Garcia$^1$, Jonathan Valamehr$^2$, and Sazadur Rahman$^1$\\
\IEEEauthorblockA{\textit{$^1$ Department of Electrical and Computer Engineering, University of Central Florida; $^2$ Intel Corporation}\\
$^1$ \{ishraq.tashdid, kimia.tasnia, alexander.garcia, mohammad.rahman\}@ucf.edu, $^2$ jonathan.k.valamehr@intel.com}}

\begin{abstract}
This work presents \emph{ATLAS}, an LLM-driven framework that bridges standardized threat modeling and property-based formal verification for System-on-Chip (SoC) security. Starting from vulnerability knowledge bases such as Common Weakness Enumeration (CWE), the framework identifies SoC-specific assets, maps relevant weaknesses, and generates assertion-based security properties and JasperGold scripts for verification. By combining asset-centric analysis with standardized threat model templates, it automates the transformation from vulnerability reasoning to formal proof. Evaluated on three HACK@DAC benchmarks, \emph{ATLAS} detected $39/48$ CWEs and, out of that $39$, \emph{ATLAS} was able to generate correct properties for $33$ of the bugs, advancing automated, knowledge-driven SoC security verification toward a secure-by-design paradigm.
\vspace{-10pt}
\end{abstract}

\keywords{Security verification, Threat Modeling, LLM for Security. \vspace{-10pt}}

\begin{document}
\maketitle
\begingroup
\renewcommand{\thefootnote}{\fnsymbol{footnote}}
\footnotetext[1]{Equal contribution.}
\endgroup
\section{Introduction}


Integrated circuits (ICs) form the backbone of modern technologies spanning artificial intelligence (AI), healthcare, autonomous systems, and defense, powering secure communication, precision-guided platforms, and advanced computing~\cite{farahmandi2023cad}. Modern ICs have evolved into highly integrated and interconnected Systems-on-Chip (SoCs) composed of diverse pre-designed intellectual property (IP) blocks. Due to the complexity, horizontally distributed business models involving untrusted supply chain entities, and the use of potentially malicious third-party IPs, every stage of SoC design can face a spectrum of security vulnerabilities~\cite{farahmandi2023cad}. Moreover, the semiconductor industry faces inherent risks from non-security-aware specifications, implementation flaws, and aggressive time-to-market demands~\cite{farahmandi2023cad}. As semiconductor systems underpin critical infrastructure, ensuring their resilience demands a proactive, secure-by-design approach. With post-silicon vulnerability remediation costs rising sharply~\cite{farzana2019soc}, early identification of threats and integration of countermeasures are essential to protect national, system-level and organizational integrity.
\begin{figure}[t]
    \centering
    \includegraphics[width=0.99\linewidth]{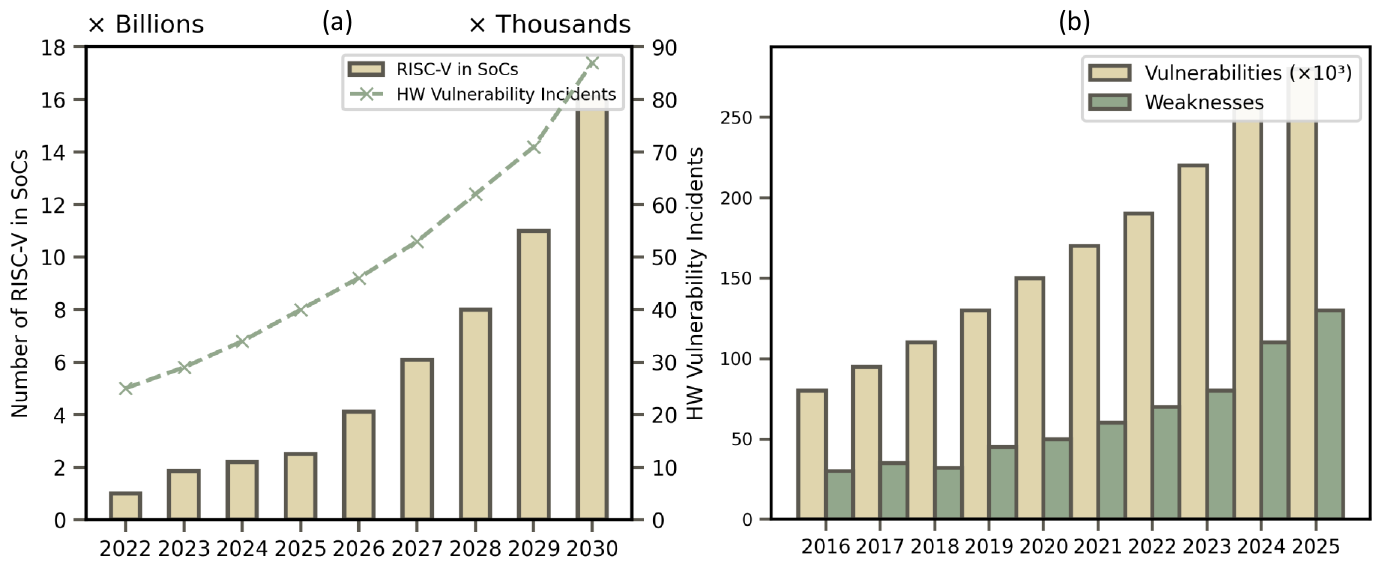}
    \vspace{-5pt}
    \caption{\small (a) Increasing hardware security vulnerabilities in RISC-V based SoCs. (b) Trend in CWE~\cite{cwe_MITRE} and CVE~\cite{cve_MITRE} over the years.}
    \vspace{-5pt}
    \label{fig:sdl}
\end{figure}

While functional verification relies on established behavioral rules from hardware descriptions, security verification remains challenging because it lacks standardized security rules and policies. As a result, much of the process depends on threat modeling. However, conventional threat modeling approaches~\cite{mcree2014microsoft,de2022threma, mitre_emb3d, tashdid2026interpuf, safesip, opl4gpt} are largely manual, error-prone, and unable to keep pace with growing design complexity and evolving attack vectors. The National Vulnerability Database~\cite{nist_nvd}, including CWE~\cite{cwe_MITRE}, CVE~\cite{cve_MITRE}, and CAPEC~\cite{barnum2008common}, catalogs over $200{,}000$ hardware and software flaws, yet current methods do not systematically use this knowledge for SoC security. Although recent efforts have leveraged these databases for threat modeling~\cite{saha2025threatlens, elsharef2024facilitating}, they still lack the comprehensiveness required for SoC security verification. Likewise, existing security verification techniques, including formal verification~\cite{rajendran2016formal,subramanyan2014formal,ahmad2022don}, code analysis~\cite{kibria2022rtl,al2023quardtropy}, hardware fuzzing~\cite{azar2022fuzz, trippel2022fuzzing, gohil2024mabfuzz, hossain2023socfuzzer}, and penetration testing~\cite{saha2025sv, ayalasomayajula2024lasp, saha2025threatlens}, remain fragmented, manual, and hard to scale. In particular, they rely on handcrafted security constraints that are labor-intensive, error-prone, and require deep expertise~\cite{kibria2024survey}. Recent LLM-assisted property generation methods are also ad hoc, covering only a small subset of vulnerabilities and lacking generalization across the full CWE space~\cite{saha2025sv,ayalasomayajula2024lasp, tasnia2025veriopt, paria2023divas}. With reported security issues increasing rapidly each year (Fig.~\ref{fig:sdl}), manual assessment is no longer feasible, making scalable and automated threat modeling and security verification essential.
To address these limitations, we present \emph{ATLAS}, a unified framework that transforms vulnerability knowledge into formal proof through contextual reasoning. 
\textbf{\ul{First,}} \emph{ATLAS} standardizes threat modeling using a comprehensive template that explicitly captures attack vectors ensuring consistency and completeness across designs. 
\textbf{\ul{Second,}} we achieve scalability and adaptability by generating a threat model database of the common security weaknesses using standard template, LLM, and publicly available vulnerability repositories CWE~\cite{cwe_MITRE}, CVE~\cite{cve_MITRE}, and CAPEC~\cite{capec} as a continuously evolving knowledge base. We have made this database open source for helping research community.
\textbf{\ul{Third,}} \emph{ATLAS} can automatically identify the SoC specifics assets and relevant threat models by analyzing generic asset definition, threat model database, and buggy SoC register transfer level (RTL) code.
\textbf{\ul{Fourth,}} through contextual reasoning with abstract syntax tree (AST), RTL summary, and SoC specification document, \emph{ATLAS} guides LLM in generating accurate, design-specific assertion-based security properties for formal verification. Evaluated on three HACK@DAC~\cite{hack@dac, hackdac18, hackdac21} benchmarks, \emph{ATLAS} identified $39/48$ relevant CWEs and produced correct properties for $33$ of those cases ($>82\%$), detecting violations and demonstrating vulnerability knowledge to formal proof.


The rest of the paper is organized as follows - Sec.~\ref{sec:background} discusses the publicly available vulnerability databases, related works in SoC security verification, and their limitations. Sec.~\ref{sec:threat_modeling} describes the asset centric threat modeling approach using standard template, adversary models, security knowledge base, and generic asset definition. Sec.~\ref{sec:sv} outlines assertion based security prorpty generation using SoC contexts and threat models from Sec.~\ref{sec:threat_modeling}. Experimental results are presented in Sec.~\ref{sec:results} before concluding the paper in Sec.~\ref{sec:conclusion}.
\vspace{-5pt}
\section{Background}\label{sec:background}
In this section we provide a brief summary of the publicly known vulnerability databases that serves as the security knowledge for \emph{ATLAS}, existing security verification methods, and their limitations.
\vspace{-15pt}
\subsection{Vulnerability Databases}\label{subsec:cwe_cve_databases}
\noindent 
To ensure SoC security, one of the first thing required is finding a comprehensive repository of real-world vulnerability incidents. The vulnerability databases maintained by the MITRE Corporation, specifically, CWE~\cite{cwe_MITRE} and CVE~\cite{cve_MITRE} represent the most comprehensive and reliable crowd-sourced repositories for security incidents across hardware and software domains. The CWE catalog defines and classifies abstract security weaknesses, such as improper privilege checks or insecure data paths, while CVEs document concrete instances of these weaknesses in specific products or platforms~\cite{mell2024hardware}. Fig.~\ref{fig:sdl}(b) shows that over the years, the number of identified CWEs and CVEs has grown significantly, reflecting both the increasing complexity of modern hardware systems and the expanding attack surface they present. For hardware security in particular, this rise underscores the urgent need for systematic, design-time mitigation strategies. By standardizing and disseminating vulnerability knowledge, these databases form a foundational resource for future security-aware SoC development.
\vspace{-5pt}
\begin{table}[b]
\caption{\small Threat model template and example.}
\centering
\label{tab:tm_template}
\resizebox{0.45\textwidth}{!}{
\setlength\tabcolsep{2pt}
\setlength\extrarowheight{0pt}
\begin{tabular}{|l|l|}
\hline
\textbf{Template} & \textbf{CWE-1245 Threat Model}\\
\hline
Adversary & Simple Hardware Adversary\\
\hline
Assets & Hardware State and Logic Integrity\\
\hline
Attack surface & Hardware Interfaces and State Machine Logic\\
\hline
Vulnerabilities & Improper or insecure FSM design \\
\hline
Threats & Exploiting undefined or insecure FSM transitions\\
\hline
\end{tabular}}
\vspace{-10pt}
\end{table}

\vspace{-5pt}
\subsection{Related works and their limitations}\label{subsec:related}
As stated previously, threat modeling is the foundation of security verification, providing a systematic way to identify and mitigate vulnerabilities before they propagate to downstream design stages. Despite its importance, existing methodologies suffer from several limitations. \circled{\small 1} Manual approaches are becoming increasingly ineffective as design complexity and attack sophistication grow~\cite{kocher2020spectre, amd_attack, intel_spectre1, intel_spectre2, fault_attack_embedded}. As attack methods evolve and the range of weaknesses expands, as shown in Fig.~\ref{fig:sdl}(a), these approaches fail to scale, limiting coverage and slowing response to new threats. \circled{\small 2} The CWE and CVE databases discussed in Sec.~\ref{subsec:cwe_cve_databases} collectively contain over $200{,}000$ hardware and software vulnerabilities, yet few techniques comprehensively leverage this knowledge to improve SoC security. \circled{\small 3} The lack of comprehensive threat modeling further reduces the effectiveness of security verification. Existing techniques can be broadly divided into static and dynamic methods. \circled{\small 4} Static methods, including formal verification~\cite{rajendran2016formal,subramanyan2014formal,ahmad2022don}, concolic testing~\cite{lyu2020scalable, lyu2019automated}, and code analysis~\cite{kibria2022rtl,al2023quardtropy}, require engineers to translate high-level security requirements into formal assertions, creating substantial manual overhead and increasing susceptibility to human error. \circled{\small 5} Dynamic methods such as fuzzing~\cite{azar2022fuzz, trippel2022fuzzing, gohil2024mabfuzz, hossain2023socfuzzer} and penetration testing~\cite{al2023sharpen, tashdid2025beyondppa} improve runtime coverage but still depend on manually crafted feedback functions, limiting automation. More importantly, current methods do not adapt well to an evolving threat landscape. Unlike area, power, and performance flows, SoC security design lacks standardized constraints or rules to guide verification, which in turn limits emerging approaches such as fuzzing and formal verification~\cite{kibria2024survey}. Recent assertion-based and large language model (LLM)-generated property methods target specific CWEs, but remain ad hoc, unscalable, and unable to generalize across the broader vulnerability space~\cite{saha2025sv, ayalasomayajula2024lasp, saha2025threatlens, tasnia2025veriopt, paria2023divas}.

\vspace{-5pt}
\section{Asset Centric Threat Modeling}\label{sec:threat_modeling}
In this section we discuss how \emph{ATLAS} performs asset centric threat modeling by defining a template and adversary models, leveraging vulnerability databases, and utilizing LLM to address the limitations of existing threat modeling methods highlighted in Sec.~\ref{subsec:related}.


\vspace{-10pt}
\subsection{Threat Model Template}\label{subsec:standard}



\begin{table}[t]
\caption{\small Adversary models in \emph{ATLAS} with different capabilities.}
\centering
\label{tab:adv}
\resizebox{0.48\textwidth}{!}{
\setlength\tabcolsep{2pt}
\begin{tabular}{|l|l|}
\hline
\textbf{Adversary Model} & \textbf{Capabilities}\\
\hline
Unprivileged Software	& Limited capability to run user-code\\
\hline
System Software & Admin/root access to system software\\
\hline
Startup code/SMM$^{1}$ & Can tamper the boot or SMM$^{1}$ code\\
\hline
Network & Can communicate with confidential services\\
\hline
Software Side Channel & Can monitor/extract confidential metadata\\
\hline
Simple Hardware & Physical access w/o expensive equipment/training\\
\hline
Skilled Hardware & Physical access with skilled training and FA$^{2}$ tools\\
\hline
Insider Threat & Anyone within the trusted supply chain\\
\hline
\multicolumn{2}{l}{\small $^1$ System management mode, $^2$ Failure analysis.}
\end{tabular}}
\end{table}

The need for a universal standard is one of the most critical limitations of existing threat modeling methods. \ul{Therefore, \emph{ATLAS} standardizes a set of elements as a template that must be defined for SoC threat modeling.} Table.~\ref{tab:tm_template} shows the threat model template elements and an example for CWE-$1245$~\cite{cwe_1245}. The example threat model is kept concise for brevity. \textit{To the best of our knowledge, this is the first attempt to standardize threat modeling}. 
\begin{itemize}[leftmargin=*]
    \item \textbf{Adversaries:} Different systems have a diverse threat landscape based on the entities involved in the design and development. Adversary models categorize different types of potential attackers by defining their scopes, capabilities, and access levels. The selected adversary model helps to decide whether a type of vulnerability can be exploited in a particular system. \emph{ATLAS} utilizes the adversary models~\cite{intel_adv} summarized in Table~\ref{tab:adv} to provide a structured framework to identify attack scenarios and vulnerabilities. 
    \item \textbf{Assets:} It is the secret data, process, or element that must be protected. Any secondary asset utilized to protect a primary asset also worth to protect. Utilizing the asset definitions in Table~\ref{tab:gen asset def}, \emph{ATLAS} performs an asset centric threat modeling in Sec.~\ref{subsec:cwe threat modeling}.
    \item \textbf{Attack Surfaces:} The adversary is expected to utilize an interface to interact with or access an asset to exploit a vulnerability. It greatly depends on the type of the system, and adversary model. For instance, an \textit{unprivileged software adversary} may leverage command prompt as an attack surface, wherever, debug interface could be more appealing for a \textit{simple hardware adversary}.
    \item \textbf{Vulnerabilities:} A vulnerability is a weakness or flaw in a system that can be exploited to compromise its security. For instance, a lack of bound checking (vulnerability) exploited by an adversary may result in a buffer overflow attack (threat).
    \item \textbf{Threats:} When an adversary exploits a vulnerability it becomes a threat. For instance, in the previous example, buffer overflow attack is the threat in this scenario. Please note that many existing threat models confuse vulnerabilities with threats~\cite{johnston2010being}. \emph{ATLAS} recognizes them as distinct elements in the threat model.
\end{itemize}

\vspace{-10pt}
\subsection{Security Knowledge Source}\label{subsec:knowledge_base}
In \emph{ATLAS}, we generate a threat model database (TMDB) that serves as the knowledge base for security verification in Sec.~\ref{sec:sv}. Several factors motivate the need for such a database.
\circled{\small 1} Functional verification has testbenches, coverage metrics, and well-defined correctness properties~\cite{farahmandi2023cad}, whereas security verification lacks comparable structure. Designers must still manually infer which threats apply to an SoC, leading to the limitations discussed in Sec.~\ref{subsec:related}.
\circled{\small 2} The National Vulnerability Database~\cite{nist_nvd} contains rich security knowledge, but it is too abstract, scattered, and unstructured for direct use in verification.
\circled{\small 3} Human-driven threat modeling cannot keep pace with SoC complexity, diverse IP blocks, and the rapidly expanding hardware attack surface, which is why existing works cover only a limited subset of CWEs~\cite{saha2025threatlens, ahmad2022don}.
\circled{\small 4} Identifying assets in RTL is only part of the task; engineers must also determine which CWEs apply to each asset class.
\circled{\small 5} Security verification~\cite{kibria2024survey} depends on accurate, repeatable mappings from weaknesses to assertion templates, making ad hoc generation error-prone and incomplete.
A threat model database standardizes and organizes vulnerability knowledge into actionable, reusable vectors, enabling automated and consistent mapping of assets to relevant weaknesses and their corresponding property families. By providing repeatable security criteria and scalable coverage across designs and evolving threats, it becomes the foundation for automated, comprehensive SoC security verification. We have published this threat model database (https://github.com/KimiaTasnia/Threat-Model-Database-for-Hardware-Security) for security professionals and researchers.
\vspace{-10pt}
\subsection{Threat Model Database Generation}\label{TMDB}
\setlength{\intextsep}{10pt}%
\setlength{\columnsep}{5pt}%
\begin{wrapfigure}{r}{0.3\textwidth}
    \centering
    \vspace{-5pt}
    \includegraphics[width=0.3\textwidth]{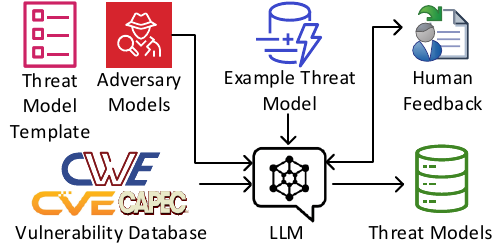}
    \caption{\small LLM assisted threat model database generation in \emph{ATLAS}.}
    \vspace{-10pt}
    \label{fig:TG Flow}
\end{wrapfigure}
Several attack patterns (CAPEC) can exploit a product vulnerability (CVE) which suffers from a fundamental weakness (CWE). For instance, both \textit{`CAPEC-1: Accessing Functionality Not Properly Constrained by ACLs'} and \textit{`CAPEC-180: Exploiting Incorrectly Configured Access Control Security Levels'} exploits the \textit{`CWE-1191: On-Chip Debug and Test Interface With Improper Access Control'} derived from CVE-2019-18827. Therefore, resistance to a specific attack pattern (e.g., CAPEC-1) does not necessarily make the developed hardware resilient against remaining attacks (e.g., CAPEC-180) exploiting the same weakness (CWE-1191). Hence, in \emph{ATLAS} we develop individual threat model for each CWE~\cite{cwe_MITRE} by following the template discussed in Sec.~\ref{subsec:standard} to generate a threat model database (TMDB) of known security weaknesses. However, we include all the associated CVEs and CAPECs in prompts to cover known vulnerabilities connected to that CWE. 
\begin{figure}[b]
    \centering
    \includegraphics[width=1\linewidth]{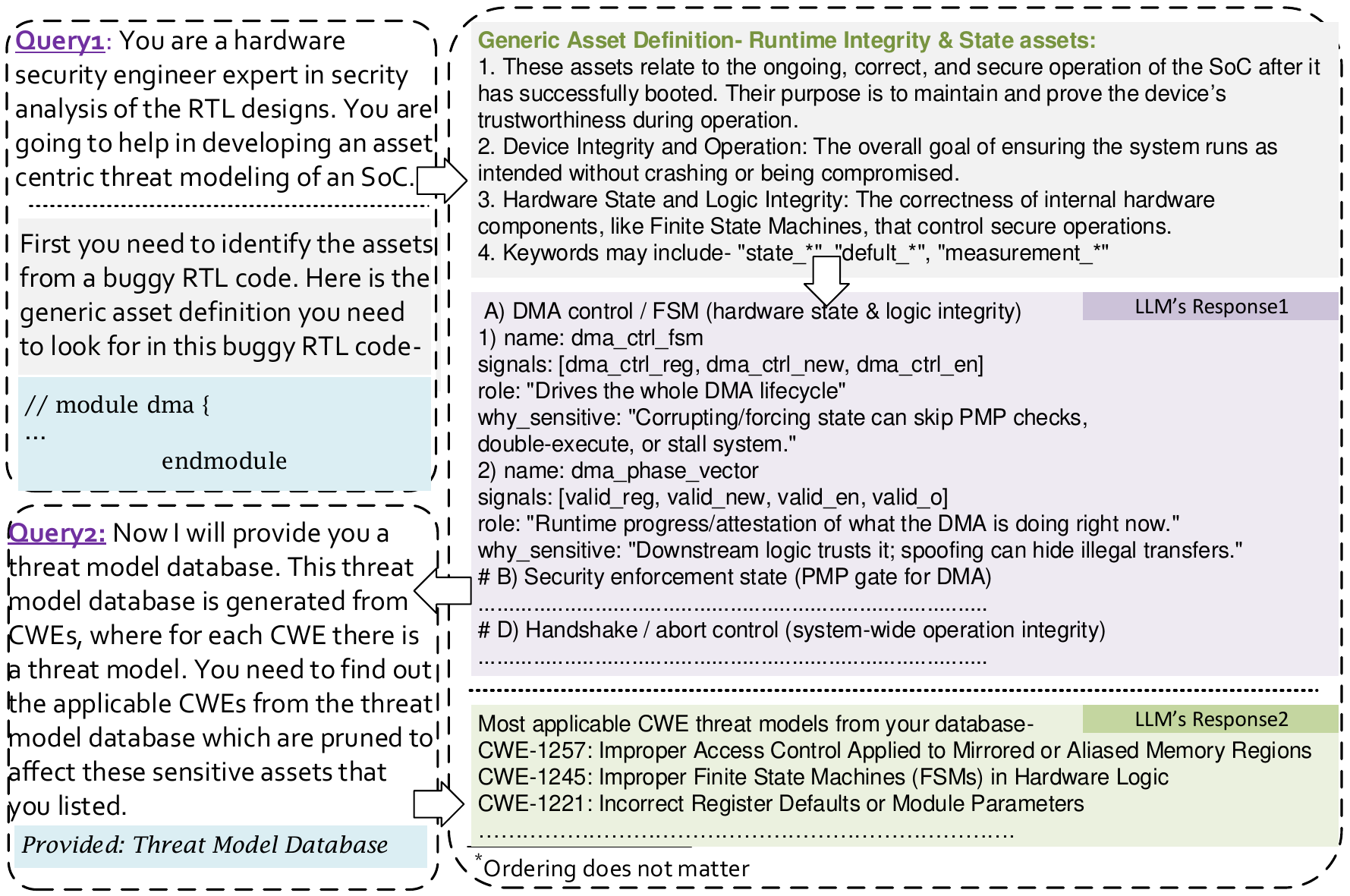}
    \caption{\small Asset-centric threat modeling from generic asset definitions (Table~\ref{tab:gen asset def}) to SoC-specific asset detection, then CWE identification using the Threat Model Database (TMDB).}
    \label{fig:Asset_CWE_TM}
\end{figure}

Fig.~\ref{fig:TG Flow} shows how we leverages GPT-5 to assist in threat model database generation using template~(Sec.~\ref{subsec:standard}), security knowledge source (Sec.~\ref{subsec:knowledge_base}), example threat models, and human feedback. To construct the threat model database (TMDB), we began by manually creating a small set of high-quality \textit{example threat models} using the standardized threat model template and adversary definitions. These examples established the expected structure and depth of analysis. GPT-5 was then prompted with these examples along with CWE~\cite{cwe_MITRE} and its associated CVE~\cite{cve_MITRE}, and CAPEC~\cite{capec} links to generate additional threat models for selected hardware CWEs, and each output was reviewed and refined through human feedback. The manually validated models were consolidated into a \textsc{JSON}-formatted dataset that included instructions, template fields, adversary descriptions, and exemplar input–output mappings. After providing GPT-5 with this context dataset, the model consistently produced complete and accurate threat models verified by human expert, enabling us to automatically generate the remaining Hardware Design CWEs and assemble a comprehensive TMDB covering all known hardware weaknesses~\cite{hardware_cwes}. We utilize this TMDB for SoC specific asset identification (Sec.~\ref{asset detection}) and threat modeling (Sec.~\ref{subsec:cwe threat modeling}).

\vspace{-10pt}
\subsection{SoC Specific Assets Detection}\label{asset detection}
\emph{ATLAS} utilizes threat model database (TMDB) and a set of generic asset definition, to identify design assets and conduct an asset centric threat modeling. Fig.~\ref{fig:Asset_CWE_TM} shows the flow of detecting SoC assets and relevant threat models for security verification in Sec.~\ref{sec:sv}. 
\vspace{-6pt}
\begin{itemize}[leftmargin=*]
    \item In an SoC, assets refer to sensitive information and data that require protection from unauthorized access and safeguarding against various threats. \emph{ATLAS} identifies these security critical assets in the RTL by providing GPT-5 a list of generic asset definitions. We categorized $7$ types of generic assets as shown in Table.~\ref{tab:gen asset def} by analyzing the NVD~\cite{cwe_MITRE, cve_MITRE, capec} that exploits various types of assets in it's database. And also the probable keywords connecting these assets in an SoC by conducting literature survey~\cite{VeriDB, HWDB}. The Table.~\ref{tab:gen asset def} is kept concise due to brevity.
    \item With the help of TMDB and generic asset definition, LLM assisted in finding the potential security assets in the given buggy SoC RTL. Fig.~\ref{fig:Asset_CWE_TM}(a) shows an example of ``Runtime Integrity \& State Resources'' with its generic definition and LLM's response to identify the associated assets in the OpenTitan Direct Memory Access (DMA) module~\cite{OpenTitan2025} utilized in several HACK@DAC competition~\cite{hack@dac}. Fig.~\ref{fig:Asset_CWE_TM}(b) is the identified assets in the dma module which lists the assets type, specific signal names, role of the signals and reasoning behind it to be asset. 
\end{itemize}

\begin{table}[t]
\caption{\small Generic assets definition for SoC specific assets detection.}
\centering
\label{tab:gen asset def}
\resizebox{\linewidth}{!}{
\setlength\tabcolsep{1pt}
\begin{tabular}{|l|l|l|}
\hline
\textbf{Asset Type} & \textbf{Definition} & \textbf{Keywords}\\
\hline
Sensitive Data &
\begin{tabular}[c]{@{}l@{}}Confidential information.\\ IP, firmware, or personal data.\end{tabular}
&
\begin{tabular}[c]{@{}l@{}}keys\_*, pass\_*, secret\_*\\ protected\_*, user\_*\end{tabular}
\\
\hline
Boot Integrity &
\begin{tabular}[c]{@{}l@{}}Verifies and loads boot.\\ Establishes hardware root of trust.\end{tabular}
&
\begin{tabular}[c]{@{}l@{}}boot\_*, hash\_*, rom\_*\\ otp\_*, key\_*\end{tabular}
\\
\hline
\begin{tabular}[c]{@{}l@{}}Attestation Data \& \\ Measurement Reports\end{tabular} &
\begin{tabular}[c]{@{}l@{}}Signed measurements of states.\\ Stored in dedicated registers.\end{tabular}
&
\begin{tabular}[c]{@{}l@{}}pcr\_*, measurement\_*\\ idev\_*\end{tabular}
\\
\hline
Parametric Data &
\begin{tabular}[c]{@{}l@{}}Device-specific non-volatile values.\\ Used for identity and calibration.\end{tabular}
&
\begin{tabular}[c]{@{}l@{}}otp\_*, trim\_*, nvm\_cfg\_*\\ uid\_*, device\_id\_*, serial\_*\end{tabular}
\\
\hline
\begin{tabular}[c]{@{}l@{}}Privileged System \\ Resources\end{tabular} &
\begin{tabular}[c]{@{}l@{}}Privileged control registers.\\ Accessible in trusted modes.\end{tabular}
&
\begin{tabular}[c]{@{}l@{}}debug\_* , *\_mode\\ *\_level\end{tabular}
\\
\hline
Shared Resources &
\begin{tabular}[c]{@{}l@{}}Shared buses or memories.\\ Arbitration to prevent leakage.\end{tabular}
&
\begin{tabular}[c]{@{}l@{}}ready\_*, valid\_*, dma\_*\\ req\_*, grant\_*, crossbar\_* \end{tabular}
\\
\hline
\begin{tabular}[c]{@{}l@{}}Runtime Integrity \& \\ State Resources\end{tabular} &
\begin{tabular}[c]{@{}l@{}}Correctness of runtime states.\\ FSM integrity and policy enforcement.\end{tabular}
&
\begin{tabular}[c]{@{}l@{}}state\_*, default\_*\\ pcr\_*, measurement\_*\end{tabular}
\\
\hline
\end{tabular}}
\end{table}

\vspace{-10pt}
\subsection{Threat Modeling with \emph{ATLAS}}
\label{subsec:cwe threat modeling}
Once ATLAS detects the SoC assets that designer must protect, it performs asset centric threat modeling at the RTL abstraction layer.

\noindent{\textbf{\underline{First,}}} we provide LLM both the SoC asset list from Sec.~\ref{asset detection} and threat model database~(TMDB) from Sec.~\ref{TMDB} as shown in Fig.~\ref{fig:ATLAS_overview}(a). The TMDB already has threat models for every hardware CWEs. \textbf{This way ATLAS covers all known hardware security threats.}

\noindent{\textbf{\underline{Second,}}} LLM runs a keyword mapping in the TMDB to match those assets in the database. For instance, for the DMA control/FSM assets in Fig.~\ref{fig:Asset_CWE_TM}(b), \emph{ATLAS} searched with the keywords (fsm, finite state machine, state, control flow, transition, deadlock) in the TMDB. It mapped these keywords in every CWE threat models and identified the most critical CWEs which are pruned to exploit those assets as shown in Fig.~\ref{fig:Asset_CWE_TM}(d). The threat model database helped tremendously in this step since it has a detailed threat model with critical components like assets, threats, vulnerabilities explaining the scenarios for any potential weakness in a system. 

\noindent{\textbf{\underline{Finally,}}} along with listing all the potential weaknesses, \emph{ATLAS} identified the corresponding threat models for the detected CWEs present in the TMDB. In this way, \emph{ATLAS} was able to generate threat models for all the possible hardware CWEs in an SoC.
\vspace{-10pt}
\section{Security Verification using Formal Proof}\label{sec:sv}
With the CWE-based SoC threat models generated in Sec.~\ref{sec:threat_modeling}, here we use formal proof to identify the vulnerabilities in a buggy SoC.


%




\vspace{-10pt}
\subsection{Rationale for Formal Verification}\label{subsec:rationale}
Security bugs are often hidden or masked in rare input combinations, resets, or corner-cases. Constraint Random Verification (CRV) covers only a small subset of those cases, and when it triggers them, it takes long time to observe the impact. On the other hand, formal verification translates the design into a formal (logic-based) model and exhaustively checks whether it satisfies given properties. If a violating execution exists, the formal tool returns a minimal counterexample trace, otherwise, it produces a proof or correctness. This aligns with our security objective of turning each SoC threat model into a property that is either formally proven or demonstrated to be vulnerable through a concrete counterexample.

Sec.~\ref{sec:threat_modeling} yields the SoC threat model with a set of relevant CWEs, however, not all are exploited in the RTL under review. To distinguish actionable vulnerabilities from non-issues, ATLAS generates an assertion based security property for each relevant CWE, capturing its corresponding attack vectors in a form suitable for formal proof. A failing property indicates that the weakness represented by that CWE is concretely exploitable in the design, whereas a proven property suggests that the RTL upholds the corresponding security requirement under the stated assumptions. 
We therefore define three objectives for formal security verification.
\noindent\textsc{\textbf{\underline{Obj1)}}} Map each relevant CWE to one or more assertion property guided by the threat model (i.e., asset, attack surface).
\noindent\textsc{\textbf{\underline{Obj2)}}} Generate concrete SystemVerilog Assertions (SVA) and formal properties bound to real signals, modes, and resets.
\noindent\textsc{\textbf{\underline{Obj3)}}} Execute the properties in JasperGold, review proofs/counterexamples, and record outcomes.

\noindent At the end of the section, we pair each objective (\textsc{\textbf{\underline{Obj}}}) with a corresponding resolution (\textsc{\textbf{\underline{Res}}}).

\begin{figure}[t]
\centering
\includegraphics[width=1.01\linewidth]{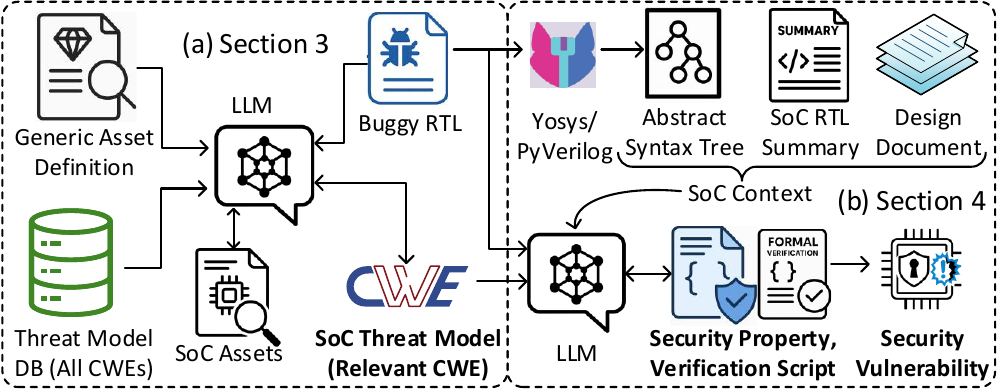}
\caption{\small Overview of AI-assisted security verification flow by \emph{ATLAS}. (a) Asset-centric threat modeling, discussed in Sec.~\ref{sec:threat_modeling}, using a generic asset definition, threat model database, and buggy SoC RTL code. The outcome of this step is SoC threat model of relevant CWEs. (b) Property-based security verification using SoC context (AST, RTL summary, and design document), threat model of the relevant CWEs.}
\label{fig:ATLAS_overview}
\end{figure}
\vspace{-5pt}
\subsection{Why SoC Context Matters}
\label{subsec:context}

Formal proofs are only as strong as the properties they verify, and those properties must be grounded in the correct design context~\cite{kibria2024survey}. While ATLAS leverages LLMs to reduce human error, minimize reliance on domain-specific expertise, and improve scalability, an SoC RTL and threat model from Sec.~\ref{sec:threat_modeling} alone cannot provide the intent required to derive sound security constraints~\cite{rogers2024securitypropertiesopensourcehardware,chang2024natural}. To produce meaningful assertions, the LLM must understand \textbf{\emph{where}} an attacker interacts (attack surfaces), \textbf{\emph{who}} controls inputs (adversary/controllability), \textbf{\emph{what}} assets must be protected, and \emph{how} the design behaves across modes, resets, and handshakes. Without these contexts, LLM-generated assertions risk being out-of-scoped, under-constrained, or misleading.

Hence, to enhance the SoC design context, we combine three \ul{complementary} sources. \textbf{\ul{First,}} the design documentation of the SoC, which contains the specifications, interface specs, register maps, mode semantics, and trust boundaries. \textbf{\ul{Second,}} the abstract syntax tree (AST) of the RTL that contains topological and structural details of the hardware (e.g., inputs, outputs, wires, registers, parameters, states, clock/reset style). And \textbf{\ul{third,}} a general summary of the SoC RTL, generated by LLM with security in mind. In the following subsections we discuss the role of each of the sources and perform an ablation study in Sec.~\ref{sec:results} (Fig.~\ref{fig:ATLAS_ablation}).
\vspace{-5pt}
\subsubsection{Design Documents (DD)} 
\label{subsubsec:doc}

Design documents are developed by architects, designers, and verification engineers and provide a rich view of intended behavior of the SoC that the base RTL cannot capture. They describe the human intended architecture, name the assets to be protected, and spell out privilege and lifecycle rules, interface semantics, and behavior~\cite{meng2024nspg}. For instance, as shown in Fig.~\ref{fig:cwe2property_flow}, once we provide the OpenTitan Direct Memory Access (DMA) documentation~\cite{OpenTitan2025} to LLM, it infers that the specification requires the DMA to enforce bounded memory ranges and Physical Memory Protection (PMP) checks and that the design must uphold a defined level of access control granularity. Additionally, this context lets us align the asset-centric CWE threat model from Sec.~\ref{sec:threat_modeling} with the design under review. One of the identified CWE in the DMA module is CWE-1190. Moreover, LLM can correlate that this is a potential violation since the module had no enable and it will trigger the start port whenever SoC is driven. Therefore, the documentation shows \textbf{\emph{where}} the attack-surface lies for the RTL.

\vspace{-5pt}
\subsubsection{AST}\label{subsubsec:ast}
As specified previously, \emph{ATLAS} provides an AST of the SoC RTL code during assertion property generation by LLM. 
Treating this structure as a graph lets the LLM reason about control and data flow~\cite{khan2025sagehlssyntaxawareastguidedllm}. In practice, this guides LLM to verify \textbf{\emph{who}} drives the signals in which conditions, trace fan-in/fan-out to policy gates, and enumerate legal state transitions before it ever writes an assertion. For example, with the DMA AST, the LLM matches CWE-1245 from SoC threat model on improper finite state machines (FSM) usage with the \textit{abort} signal (See Fig.~\ref{fig:cwe2property_flow}.(b)). Hence, the AST guides LLM to flag that \textit{abort} could persist or be \textit{X} in paths that should be clean when \textit{done\_i} is low. This matches a known vulnerability that was actually exploited in the HACK@DAC 21~\cite{hackdac21}. 

\vspace{-5pt}
\subsubsection{RTL Summary}
\label{subsubsec:summary}

\begin{figure}[t]
  \centering
  \includegraphics[width=1.01\linewidth]{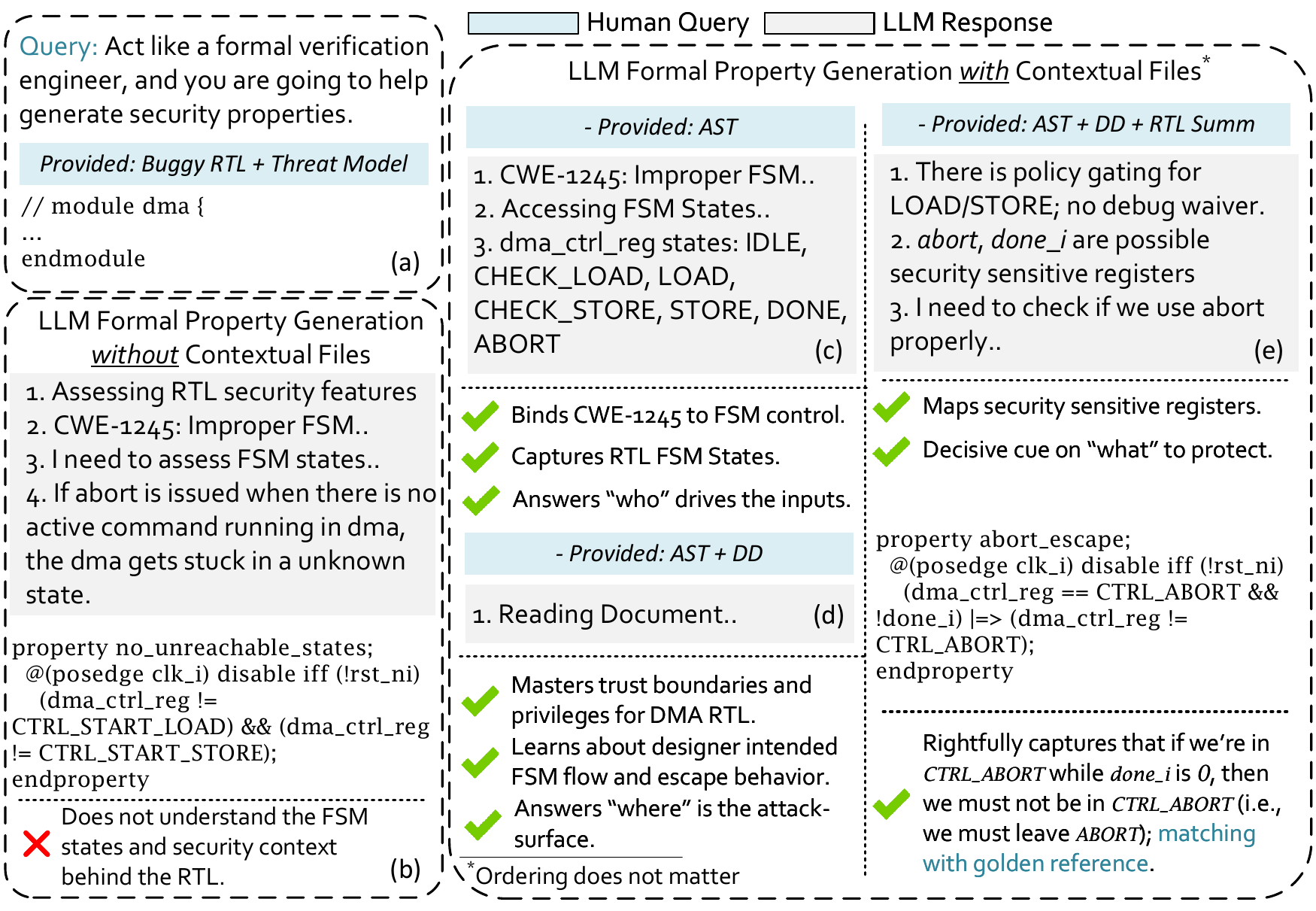}
  \vspace{-15pt}
  \caption{\small LLM–assisted security property generation. 
  (a) The same prompt (buggy SoC RTL and Threat Model) is given to both methods. (b) Plain LLM: without context, LLM misreads the design intent and fails to derive the correct security property. (c) \emph{ATLAS}: adding AST provides control and structural knowledge of the FSM behavior. (d) The design documents (DD) then highlight the trust boundaries and attack-surface. (e) RTL summary exposes the sensitive registers. Taken together, \emph{ATLAS} is able to generate the correct property.}
  \label{fig:cwe2property_flow}
\end{figure}

The RTL summary is a security-centric synopsis produced by the LLM directly from the source code. It gives LLM an objective perspective during property generation by highlighting security critical assets, guardrails, rules and checks without relying on human hints. This reduces the bias, a concept similar to why we do CRV instead of directed verification. The summary also surfaces code-level nuances that are easy to miss in a non-security-centric design document. Revisiting the DMA example from Fig.~\ref{fig:cwe2property_flow}, LLM is able to note \textbf{\emph{what}} are the critical registers and assets that are vulnerable, (not explicitly stated in the design documentation).

\begin{table*}[t]
\centering
\scriptsize
\setlength\tabcolsep{2pt}
\caption{\centering \small Security verification by \emph{ATLAS} across three HACK@DAC~\cite{hack@dac} benchmarks by detecting relevant CWE threat models, generating corresponding assertion based security properties, and performing verification by JasperGold.}
\label{tab:eval}
\resizebox{1.01\textwidth}{!}{
\begin{tabular}{@{} l l c c c | l l c c c | l l c c c @{}}
\toprule
\multicolumn{5}{c}{\textbf{HACK@DAC18~\cite{hackdac18}}} &
\multicolumn{5}{c}{\textbf{HACK@DAC19~\cite{hackdac19}}} &
\multicolumn{5}{c}{\textbf{HACK@DAC21~\cite{hackdac21}}} \\
\cmidrule(lr){1-5}\cmidrule(lr){6-10}\cmidrule(lr){11-15}
\textbf{Module} & \textbf{Bugs} & \textbf{Detected CWEs} & \textbf{Prop.} & \textbf{FV}
& \textbf{Module} & \textbf{Bugs} & \textbf{Detected CWEs} & \textbf{Prop.} & \textbf{FV}
& \textbf{Module} & \textbf{Bugs} & \textbf{Detected CWEs} & \textbf{Prop.} & \textbf{FV} \\

\cmidrule(lr){1-1}\cmidrule(lr){2-2}\cmidrule(lr){3-3}\cmidrule(lr){4-4}\cmidrule(lr){5-5}\cmidrule(lr){6-6}\cmidrule(lr){7-7}\cmidrule(lr){8-8}\cmidrule(lr){9-9}\cmidrule(lr){10-10}\cmidrule(lr){11-11}\cmidrule(lr){12-12}\cmidrule(lr){13-13}\cmidrule(lr){14-14}\cmidrule(lr){15-15}
adbg\_tap
& \begin{tabular}[t]{@{}l@{}}Logic bug\\Checks 31/32 bits\\Pwd not reset\\Reset broken\end{tabular}
& \begin{tabular}[t]{@{}l@{}}\detcwe{1221}, 1191, 1271\\\detcwe{1298}, 1221, 1191\\\detcwe{1329}, 1271, 276\\\st{\textbf{1419}}, 1271, 1191\end{tabular}
& \begin{tabular}[t]{@{}l@{}}\checkmark\\\checkmark\\\checkmark\\\xmark\end{tabular}
& \begin{tabular}[t]{@{}l@{}}\checkmark\\\checkmark\\\checkmark\\\xmark\end{tabular}
& csr\_regfile
& \begin{tabular}[t]{@{}l@{}}Debug exit issue\\Read/Write access\\Incorrect M-mode\\Incorrect update\end{tabular}
& \begin{tabular}[t]{@{}l@{}}\detcwe{1244}, 1191, 1243\\\detcwe{{\color{blue}{1262}}}, 1220, 269\\\detcwe{1262}, 1220, 269\\\detcwe{{\color{blue}{1256}}}, 200, 203\end{tabular}
& \begin{tabular}[t]{@{}l@{}}\xmark\\\checkmark\\\checkmark\\\checkmark\end{tabular}
& \begin{tabular}[t]{@{}l@{}}\xmark\\\checkmark\\\checkmark\\\checkmark\end{tabular}
& dmi\_jtag
& \begin{tabular}[t]{@{}l@{}}Pass flag reset\\Write w/o pass\\Hardcoded key\\Unreachable state\end{tabular}
& \begin{tabular}[t]{@{}l@{}}\st{\textbf{1239}}, 1271, 1191\\\detcwe{1245}, 1191, 1244\\\detcwe{1329}, 321, 798\\\detcwe{1245}, 1191, 1221\end{tabular}
& \begin{tabular}[t]{@{}l@{}}\xmark\\\checkmark\\\checkmark\\\checkmark\end{tabular}
& \begin{tabular}[t]{@{}l@{}}\xmark\\\checkmark\\\checkmark\\\checkmark\end{tabular}
\\
\cmidrule(lr){1-5}\cmidrule(lr){6-10}\cmidrule(lr){11-15}
mux\_func
& \begin{tabular}[t]{@{}l@{}}Wrong mux\\Out uncleared\end{tabular}
& \begin{tabular}[t]{@{}l@{}}\detcwe{1240}, 1245, 1221\\\detcwe{325}, 226, 1271\end{tabular}
& \begin{tabular}[t]{@{}l@{}}\checkmark\\\checkmark\end{tabular}
& \begin{tabular}[t]{@{}l@{}}\checkmark\\\checkmark\end{tabular}
& commit\_stage
& \begin{tabular}[t]{@{}l@{}}Interrupt\\ $\times2$ Commit\end{tabular}
& \begin{tabular}[t]{@{}l@{}}\detcwe{1281}, 1262, 1220\\\detcwe{{\color{blue}{1281}}}, 1241, 1271\end{tabular}
& \begin{tabular}[t]{@{}l@{}}\xmark\\\xmark\end{tabular}
& \begin{tabular}[t]{@{}l@{}}\xmark\\\xmark\end{tabular}
& csr\_regfile
& \begin{tabular}[t]{@{}l@{}}Debug open\\Privilege leak\end{tabular}
& \begin{tabular}[t]{@{}l@{}}\detcwe{1220}, 1191, 1243\\\detcwe{1262}, 1220, 269\end{tabular}
& \begin{tabular}[t]{@{}l@{}}\checkmark\\\checkmark\end{tabular}
& \begin{tabular}[t]{@{}l@{}}\checkmark\\\checkmark\end{tabular}

\\
\cmidrule(lr){1-5}\cmidrule(lr){6-10}\cmidrule(lr){11-15}
riscv\_cs\_reg
& Wrong privilege
& \begin{tabular}[t]{@{}l@{}}\detcwe{1207}, 1262, 269\end{tabular}
& \checkmark
& \checkmark
& axi\_node
& Improper bypass
& \begin{tabular}[t]{@{}l@{}}\detcwe{1220}, 1191, 269\end{tabular}
& \xmark
& \xmark
& aes\_192
& Counter stuck
& \begin{tabular}[t]{@{}l@{}}\detcwe{1240}, 323, 330\end{tabular}
& \checkmark
& \checkmark
\\
\cmidrule(lr){1-5}\cmidrule(lr){6-10}\cmidrule(lr){11-15}
apb\_gpio
& \begin{tabular}[t]{@{}l@{}}Lock writable\\Reset clears lock\end{tabular}
& \begin{tabular}[t]{@{}l@{}}\detcwe{1207}, 1220, 276\\\detcwe{1206}, 276, 1271\end{tabular}
& \begin{tabular}[t]{@{}l@{}}\xmark\\\checkmark\end{tabular}
& \begin{tabular}[t]{@{}l@{}}\xmark\\\checkmark\end{tabular}
& ariane
& \begin{tabular}[t]{@{}l@{}}No flush after atomic\\No flush after priv \end{tabular}
& \begin{tabular}[t]{@{}l@{}}\detcwe{{\color{blue}{1281}}}, 1206, 1271\\\detcwe{{\color{blue}{1281}}}, 1191, 1241\end{tabular}
& \begin{tabular}[t]{@{}l@{}}\checkmark\\\checkmark\end{tabular}
& \begin{tabular}[t]{@{}l@{}}\checkmark\\\checkmark\end{tabular}
& reglk\_wrapp.
& \begin{tabular}[t]{@{}l@{}}Locks disabled\\Locks dropped\end{tabular}
& \begin{tabular}[t]{@{}l@{}}\st{\textbf{1232}}, 1221, 276\\\st{\textbf{1234}}, 1232, 1221\end{tabular}
& \begin{tabular}[t]{@{}l@{}}\xmark\\\xmark\end{tabular}
& \begin{tabular}[t]{@{}l@{}}\xmark\\\xmark\end{tabular}
\\
\cmidrule(lr){1-5}\cmidrule(lr){6-10}\cmidrule(lr){11-15}
axi\_address\_dec.
& Errors ignored
& \begin{tabular}[t]{@{}l@{}}\st{\textbf{20}}, 703, 388\end{tabular}
& \xmark
& \xmark
& controller
& Halt w/o exception
& \begin{tabular}[t]{@{}l@{}}\detcwe{1207}, 1419, 1271\end{tabular}
& \checkmark
& \xmark
& sha256\_wrapp.
& Input uncleared
& \begin{tabular}[t]{@{}l@{}}\st{\textbf{1239}}, 226, 1271\end{tabular}
& \xmark
& \xmark
\\
\cmidrule(lr){1-5}\cmidrule(lr){6-10}\cmidrule(lr){11-15}
periph\_bus
& \begin{tabular}[t]{@{}l@{}}Dual overlap\\GPIO range alias\\Triple overlap\end{tabular}
& \begin{tabular}[t]{@{}l@{}}\detcwe{1260}, 1203, 1257\\\detcwe{1257}, 1203, 1260\\\detcwe{1260}, 1257, 1203\end{tabular}
& \begin{tabular}[t]{@{}l@{}}\checkmark\\\checkmark\\\checkmark\end{tabular}
& \begin{tabular}[t]{@{}l@{}}\checkmark\\\checkmark\\\checkmark\end{tabular}
&  &  & &
&
& aes0\_wrapp.
& \begin{tabular}[t]{@{}l@{}}Input uncleared\\Ungated debug\\Uncleared debug\end{tabular}
& \begin{tabular}[t]{@{}l@{}}\detcwe{226}, 1243, 1271\\\detcwe{1243}, 1191, 1243\\\st{\textbf{1258}}, 1243, 226\end{tabular}
& \begin{tabular}[t]{@{}l@{}}\checkmark\\\checkmark\\\xmark\end{tabular}
& \begin{tabular}[t]{@{}l@{}}\checkmark\\\checkmark\\\xmark\end{tabular}
\\
\cmidrule(lr){1-5}\cmidrule(lr){6-10}\cmidrule(lr){11-15}
riscv\_core
& \begin{tabular}[t]{@{}l@{}}Halted access\\FSM hanged\\Incomplete case\\Insecure req\end{tabular}
& \begin{tabular}[t]{@{}l@{}}\st{\textbf{1298}}, 1191, 1262\\\detcwe{1245}, 1241, 1271\\\detcwe{1245}, 1241, 1271\\\detcwe{1220}, 1262, 269\end{tabular}
& \begin{tabular}[t]{@{}l@{}}\xmark\\\checkmark\\\checkmark\\\checkmark\end{tabular}
& \begin{tabular}[t]{@{}l@{}}\xmark\\\checkmark\\\checkmark\\\checkmark\end{tabular}
& \atlaslegendcell
& acct\_wrapp.
& Reset full access
& \begin{tabular}[t]{@{}l@{}}\detcwe{276}, 1221, 1220\end{tabular}
& \xmark
& \xmark
\\
\cmidrule(lr){1-5}\cmidrule(lr){11-15}
rtc\_clock
& RTC time bug
& \begin{tabular}[t]{@{}l@{}}\detcwe{1247}, 682, 1271\end{tabular}
& \checkmark
& \checkmark
& \atlaslegendempty
& dma
& Abort sticky
& \begin{tabular}[t]{@{}l@{}}\detcwe{1245}, 1271, 1221\end{tabular}
& \checkmark
& \checkmark
\\

\cmidrule(lr){1-5}\cmidrule(lr){11-15}
soc\_intercon.
& Out-of-range
& \begin{tabular}[t]{@{}l@{}}\detcwe{1203}, 1260, 1257\end{tabular}
& \xmark
& \xmark
& \atlaslegendempty
& rsa\_wrapp.
& Output uncleared
& \begin{tabular}[t]{@{}l@{}}\detcwe{226}, 1271, 1271\end{tabular}
& \checkmark
& \checkmark
\\

\cmidrule(lr){1-5}\cmidrule(lr){11-15}
jtag\_tap\_top
& No password
& \begin{tabular}[t]{@{}l@{}}\detcwe{1262}, 1191, 1220\end{tabular}
& \checkmark
& \xmark
& \atlaslegendempty
& riscv\_periph.
& Hardcoded ROM
& \begin{tabular}[t]{@{}l@{}}\st{\textbf{1310}}, 1323, 1312\end{tabular}
& \xmark
& \xmark
\\
\bottomrule
\end{tabular}
}
\vspace{-5pt}
\end{table*}
\vspace{-10pt}

\subsection{CWE to Property Generation}
\label{subsec:property_generation}

After obtaining the CWE-based SoC threat model from Sec.~\ref{sec:threat_modeling}, we map it to assertion-based security properties using the contexts discussed in Sec.~\ref{subsec:context}. Fig.~\ref{fig:cwe2property_flow} compares two pathways that start from the same prompt (Fig.~\ref{fig:cwe2property_flow}(a)). In (b), the LLM uses \textit{only} buggy SoC RTL and the TMDB and fails to generate the correct property. In contrast, under the \emph{ATLAS} flow, when the same buggy SoC RTL and threat model from (a) are augmented with the additional contexts ((c) AST, (d) Design Documents, and (e) RTL summary), the LLM produces the correct security property in almost every case, and these map cleanly to formal proofs (Sec.~\ref{subsubsec:hack@dac_results}). \textsc{\textbf{\underline{Res1)}}} For CWE-1245 in Fig.~\ref{fig:cwe2property_flow}, the LLM first reviews the general system behavior, register map, RTL scope, and I/O connectivity from the DD. The AST then reveals where policy/privilege gating, reset/initialization hygiene, FSM integrity, ordering, and control flow are implemented, including RTL details not explicit in the DD. Next, a security-centric RTL summary helps bind assets, policies, and states, and provides the key cue to condition \emph{ATLAS} on the adversary model. Using all three sources, the LLM performs an initial pruning \{\( \text{asset} (\hyperref[subsubsec:summary]{{\mathrm{Summary}}})  \cap \text{attack-surface} (\hyperref[subsubsec:doc]{{\mathrm{DD}}}) \cap \text{path} (\hyperref[subsubsec:ast]{{\mathrm{AST}}}) \)\}, or at minimum forms concrete hypotheses about how sensitive assets should behave in the RTL. Our experiments also showed that the ordering of these contexts has little to no effect. \textsc{\textbf{\underline{Res2)}}} With the detailed signal names, state encodings, logical semantics, and interface rules, \emph{ATLAS} then instantiates non-vacuity covers so antecedents are exercised, providing guideposts for SVA generation. \textsc{\textbf{\underline{Res3)}}} Finally, the resulting properties act as design-rule checks for JasperGold~\cite{Cadence_JasperGold}: the properties are emitted as assertion files together with a \textit{.tcl} harness that compiles the DUT, applies constraints, and runs proofs. This yields the outcome targeted in Fig.~\ref{fig:ATLAS_overview}. Inconclusive results can be refined by tightening or relaxing assumptions, adding clarification, and rerunning twice. Persistently inconclusive cases are flagged for manual review.

\vspace{-5pt}
\section{Results \& Evaluation}\label{sec:results}

In this section, we evaluate \emph{ATLAS} on three industry scale RISC-V based SoC designs. We report detection coverage, properties, and formal proof outcomes, compare against a manual proof baseline~\cite{rogers2024securitypropertiesopensourcehardware} as golden reference, and then ablate the three contexts.
\vspace{-20pt}
\subsection{Experimental Setting}
\label{subsec:exp_setting}
OpenTitan SoC~\cite{openTitan} featured in the HACK@DAC benchmarks (’18, ’19, ’21)~\cite{hackdac18, hackdac19, hackdac21} offers a rich set of real-world security bugs, making it an ideal benchmark for assessing ATLAS.
\emph{ATLAS} is assisted by GPT-5 as the base LLM. Design documents for the buggy RTL are sourced from the OpenTitan documentation and repository~\cite{OpenTitan2025}. ASTs are generated using Yosys~\cite{wolf2013yosys} or PyVerilog with Slang for SystemVerilog support. RTL summary is generated by the LLM to minimize human bias. JasperGold is used for formal verification.

\begin{figure}[b]
    \centering
    \includegraphics[width=\linewidth]{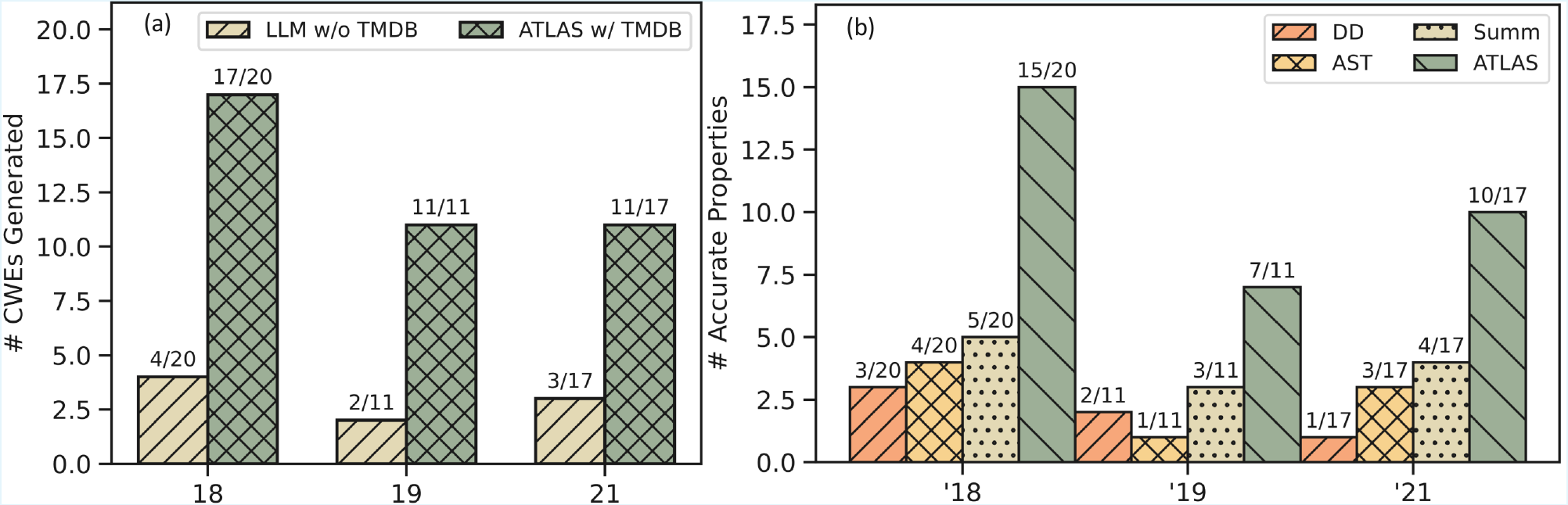}
    \caption{\small
    TMDB ablation across HACK@DAC'18, '19, and '21. (a) Number of correct CWEs identified by an LLM without TMDB versus \emph{ATLAS} with TMDB.
    (b) Context ablation on accurate properties.}
    \label{fig:ATLAS_ablation}
\end{figure}
\vspace{-10pt}
\subsection{Evaluation}
\label{subsec:eval}
\subsubsection{Benchmark Results}
\label{subsubsec:hack@dac_results}

\emph{ATLAS} runs automatically on each benchmark in Table~\ref{tab:eval}. To keep the process fast and predictable, we limit the search to at most three iterations per bug; unresolved cases are flagged for manual review. Most failures within this budget follow a common pattern: either the bug depends on subtle functional intent not stated in the public documentation, or the specification omits the exact corner case the property must capture. Across the three HACK@DAC~\cite{hack@dac} benchmarks, \emph{ATLAS} shows strong performance. The Detected CWEs column in Table~\ref{tab:eval} lists the CWE-based SoC threat models generated by \emph{ATLAS} as discussed in Sec.~\ref{sec:threat_modeling}. For each bug, \emph{ATLAS} often identifies multiple relevant CWEs. During security verification (Sec.~\ref{sec:sv}), it then detects the correct exploited CWE (shown in bold in Table~\ref{tab:eval}) by generating assertion-based security properties and checking them in JasperGold for formal proof.

A checkmark (\checkmark) in the property (Prop.) and formal verification (FV) columns of Table~\ref{tab:eval} indicates that \emph{ATLAS} successfully generated the assertion-based security property and detected the weakness in the given HACK@DAC~\cite{hack@dac} module. \textbf{With our threat model database (TMDB), \emph{ATLAS} detected $39/48$ security weaknesses (CWEs) across the three HACK@DAC~\cite{hack@dac} benchmarks. More importantly, when the correct CWE was identified, \emph{ATLAS} generated the correct property in more than $82\%$ of the cases.} When \emph{ATLAS} does not detect the exact CWE, it still produces \hyperref[subsubsec:relevant_cwes]{{\textit{“relevant”}}} CWEs from the SoC threat model in Sec.~\ref{sec:threat_modeling}. Even when formal verification fails, the generated property is often still useful. Such failures often occur because the harness does not exercise the exact exploit path or because internal guards mask the issue under nominal stimuli. In these cases, the property can still point to a real SoC weakness. For example, several control and privilege checks generated by \emph{ATLAS} were not proven in the given environment, yet they revealed missing controls or incomplete resets that require fixes, consistent with the golden reference~\cite{rogers2024securitypropertiesopensourcehardware}. We treat these cases as actionable warnings rather than false alarms. When \emph{ATLAS} succeeds, high-quality context greatly improves security verification: contexts with clear interface and state descriptions yield higher first-pass success rates. This is visible in Table~\ref{tab:eval}, where \emph{ATLAS} consistently detected the correct CWE and generated the appropriate property for well-documented blocks such as \textit{aes\_192} and \textit{rsa\_wrapper}. By contrast, missed CWE detections usually correspond to underspecified designs. For example, in HACK@DAC’18~\cite{hackdac18}, the first \textit{adbg\_tap} bug is described only as a generic logic error, and the documentation does not specify the intended latch and mask behavior. Similarly, in HACK@DAC’21~\cite{hackdac21}, the reset behavior of the \textit{dmi\_jtag} pass flag is not specified. In both cases, the lack of explicit functional contracts makes automatic property construction difficult for \emph{ATLAS}.

To understand deeper into how contexts help, in the following subsection, we show how the TMDB and three SoC contexts remain essential to steer the LLM towards meaningful assertions.
\subsubsection{Ablation Study}\label{subsubsec:ablation}
We quantify the effect of \emph{TMDB} and each contexts (DD, AST, RTL Summ) across HACK@DAC'18, '19, and '21. Looking at Fig.~\ref{fig:ATLAS_ablation}(a), removing TMDB completely collapses threat modeling and security verification. For CWE detection, an LLM‐only baseline detects $6/20$ on ’18, $3/11$ on ’19, and $4/17$ on ’21, which is less than $30\%$ detection rate. In contrast, \emph{ATLAS} identifies $17/20$, $11/11$, and $11/17$, respectively, with the TMDB, highlighting how structured threat modeling can improve LLM performance.

Moreover, what makes \emph{ATLAS} capable lies in the three SoC contexts. Ablation on Fig.~\ref{fig:ATLAS_ablation}(b) shows the same trend. When provided only one SoC context, formal verification detects $\leq5$ accurate properties. \emph{ATLAS} detects $+10$, $+4$, and $+6$ additional correct properties in HACK@DAC18, 19, and 21 benchmarks corresponding to roughly $3\times$, $2.3\times$, and $2.5\times$ improvement. It is appearant that, TMDB is the root of identifying the CWEs, which then, with the help of three SoC contexts, results to accurate property generation. 
\vspace{-16pt}
\subsubsection{Relevant CWE and Property Generation}\label{subsubsec:relevant_cwes}
During CWE generation from TMDB, \emph{ATLAS} often lists more than one \emph{relevant} CWE because the same asset can have multiple weaknesses. One may be truly exploited while a nearby rule on that asset still holds. In HACK@DAC’18 \textit{apb\_gpio}, a real weakness exists in the lock register where the key can be overwritten. \emph{ATLAS} also proposed a related property for the same lock-register asset that checks whether the address decoding for the key is correct. Furthermore, our experiments showed strong correlation among CWEs. A single weakness can overlap with two CWEs that describe nearly the same rule using different terms. For instance, in \textit{csr\_regfile} from HACK@DAC’21, \emph{ATLAS} detected CWE-1220 and CWE-1262 together for the same debug scenario. One emphasizes access control policy, while the other focuses on privilege enforcement. In practice, both capture the idea that entering debug must not weaken privilege. Hence, \emph{ATLAS} reports both CWEs together.
\vspace{-10pt}
\section{Conclusion and Future Directions}\label{sec:conclusion}
ATLAS bridges standardized threat modeling and property-based formal verification for SoC security. It builds a CWE-driven threat model database and combines design documents, AST, and RTL summaries to generate design-specific security assertions end-to-end, from asset identification to weakness mapping to formal proof. Across three HACK@DAC benchmark suites, ATLAS reliably identifies vulnerabilities and produces correct formal properties with minimal human intervention, outperforming prior approaches and showing meaningful progress in SoC design and security verification.
\newpage
\bibliographystyle{ACM-Reference-Format}
\bibliography{refs}

\end{document}